\documentclass[preprint,pteplogo]{ptephy_v2}

\preprintnumber{2606.01699} 
\usepackage{hyperref}

\usepackage{amssymb, amsfonts, amsmath, mathtools, bm}
\usepackage{color}
\usepackage{booktabs}
\usepackage[caption=false]{subfig}
\usepackage{mathrsfs}
\usepackage{enumerate}
\usepackage{amsfonts}
\usepackage{graphicx}
\usepackage{enumitem}
\usepackage{setspace}
\usepackage{bm}
\usepackage{url}

\usepackage[
]{hyperref}
\hypersetup{%
 setpagesize=false,
 bookmarksnumbered=true,%
 bookmarksopen=true,%
 colorlinks=true,%
 linkcolor=blue,%
 citecolor=blue}

\newcommand{\rmd}{\mathrm{d}}
\newcommand{\bra}[1]{\left( #1 \right)}
\newcommand{\brb}[1]{\left[ #1 \right]}

\newcommand{\miki}[1]{{\color{cyan}{#1}}}

\begin{document}

\title{Ellis-Bronnikov Wormhole Shadows with Spherically Symmetric Accretion Flow}


\author[1]{Mikiya M. Takahashi}
\affil[1]{Department of Liberal Arts, National Institute of Technology, Tokyo College, 1220-2, Kunugida-machi, Hachioji, Tokyo, 193-0942, Japan \email{m\_takahashi@tokyo-ct.ac.jp}}

\author[2,3]{Keisuke Nakashi}
\affil[2]{Kaichi Tokorozawa Secondary School, 169 Ohaza Matsugo, Tokorozawa City, 
Saitama 359-0027, Japan}
\affil[3]{Department of Physics, Rikkyo University, Toshima, Tokyo 171-8501, Japan}

\begin{abstract}%
We investigate the observational differences between the Ellis--Bronnikov (EB) wormhole and the Schwarzschild black hole (BH) by performing general relativistic radiative transfer (GRRT) simulations. 
We consider a spherically symmetric steady-state accretion flow and perform GRRT simulations incorporating synchrotron emission. 
For both the EB wormhole and the Schwarzschild BH, the simulated images consist of a central shadow region and a bright photon ring. 
We find that both the shadow region and the photon ring of the EB wormhole are brighter than those of the Schwarzschild BH. 
These differences arise from the absence of an event horizon in the EB wormhole, allowing the emission from the accreting matter around and beyond the throat to contribute to the intensity. 
We also compare the simulated images with the Event Horizon Telescope (EHT)
observations of M87*. 
Although the total flux and the ring diameter are comparable with the EHT
results, our models with the spherically symmetric accretion flow cannot reproduce the deep central brightness depression observed by the EHT, for both the EB wormhole and the Schwarzschild BH. 
This indicates that a non-spherical accretion flow is required for a detailed comparison with the EHT image.
\end{abstract}

\subjectindex{E00, E30, E34}

\maketitle

\section{Introduction}
Images of M87* and Sgr A* have been reported by the Event Horizon Telescope (EHT) Collaboration~\cite{EHT1,EHT_Sgr1,EHT6,EHT_Sgr6}. 
The bright ring structures and associated shadows in the EHT images are produced by photons coming from the vicinity of the central compact object. 
Thus, the EHT images provide an important probe of the nature of strong gravitational fields around compact objects.
The EHT data suggest that the central compact object is a Kerr black hole. 
However, the possibility that the central object is not a Kerr black hole has been actively discussed.
The EHT Collaboration calculated images and broadband spectra in various spacetimes to constrain deviations from the Kerr metric \citep{EHT_Sgr6}, and concluded that compact objects with a thermal surface can be ruled out. 
On the other hand, the image size alone constrains deviations from the Kerr metric only at the $\sim 10\%$ level~\cite{EHT_Sgr6}. 
This suggests that alternative models of the central compact object have not yet been fully ruled out by the current EHT observations.
In fact, realistic images of a black hole with dilaton hair have been calculated using general relativistic magnetohydrodynamics (GRMHD) and general relativistic radiative transfer (GRRT) simulations~\cite{Mizuno2018, Roder2023}. 
As a result, the authors pointed out that it is difficult to distinguish between the Kerr black hole and a dilatonic black hole from the images with the current EHT accuracy. 
Furthermore, it has also been pointed out that time variations in the radio flux and radio interferometry with extremely long baselines are useful for identifying the nature of compact objects \citep{Moriyama2025, Suzuki2025}.

Beyond black holes with non-Kerr metrics, a variety of exotic compact objects have been proposed as alternative candidates that can produce shadow images remarkably similar to those of black holes~\cite{Vagnozzi:2022moj}.
Among such candidates, wormholes have attracted attention as black hole mimickers.
A wormhole is a horizonless compact object with a throat connecting two distant regions.
In general relativity (GR), exotic matter that violates the null energy condition is required to sustain the wormhole structure~\cite{Morris:1988cz, Morris:1988tu}.
In this paper, we consider the Ellis--Bronnikov (EB) wormhole~\cite{Ellis1973,Bronnikov:1973fh}, which is a traversable wormhole solution in GR supported by a massless phantom scalar field.
The stability of the EB wormhole has been studied extensively.
It is known that the EB wormhole is unstable against radial linear perturbations~\cite{Shinkai:2002gv,Gonzalez:2008wd,Gonzalez:2008xk,Bronnikov:2011if,Bronnikov:2012ch,Xu:2025jad}.
However, it has been shown that the instability can be mitigated by introducing additional matter fields~\cite{Das:2005un,Shatskiy:2008us,Bronnikov:2013coa} or by considering modified theories of gravity~\cite{Kanti:2011jz,Kanti:2011yv}. 
The quasinormal modes of the EB wormhole have been also investigated~\cite{Blazquez-Salcedo:2018ipc,Azad:2022qqn,Azad:2023idq,Khoo:2024yeh}. 

Numerous previous studies have investigated the observational signatures of wormholes.
In particular, their optical signatures have attracted considerable attention as a means of distinguishing wormholes from black holes.
Gravitational lensing provides a powerful tool to probe the wormhole geometry. 
Gravitational lensing properties of the EB wormhole have been investigated, including gravitational microlensing and light curves~\cite{Chetouani:1984qdm,Cramer:1994qj,Torres:1998xd,Safonova:2001vz,Bogdanov:2008zy,Abe:2010ap,Toki:2011zu,Kitamura:2012zy,Tsukamoto:2016zdu}, Einstein rings~\cite{Tsukamoto:2012xs}, and deflection angles in both the weak and strong deflection regimes~\cite{Clement:1982ej,Nandi:2006ds,Dey:2008kn,Muller:2008zza,Bhattacharya:2010zzb,Gibbons:2011rh,Nakajima:2012pu,Tsukamoto:2012np,Tsukamoto:2016qro,Tsukamoto:2016jzh,Tsukamoto:2017edq,Jusufi:2017gyu,Cai:2023ite}. 
More recently, motivated by the EHT observations, attention has shifted toward the direct imaging of wormholes, including their shadows and the images formed by surrounding accretion flows~\cite{Bambi:2013nla,Nedkova:2013msa,Ohgami:2015nra,Ohgami:2016iqm,Tsukamoto:2017hva,Shaikh:2018kfv,Paul:2019trt,Bambi:2019tjh,Bugaev:2021dna,Bambi:2021qfo,Ishkaeva2023,Huang:2023yqd,Alloqulov:2024olb,Novikov:2025elo,Xia2026,Ishkaeva:2026tot}. 

The observational appearance of the central compact object depends strongly on the properties of the accretion flow surrounding it, such as the geometry and optical depth of the accretion flow. 
In particular, Ohgami and Sakai~\cite{Ohgami:2015nra} computed shadow images of the massless EB wormhole illuminated by a steady-state spherical dust flow. 
In the present paper, we extend their work to the EB wormhole and to more general spherically symmetric accretion flows. 
Unlike many previous studies in which the profiles of the accretion flow were prescribed by hand, we self-consistently derive the steady-state spherical flow solution in the EB wormhole spacetime. 
Furthermore, we calculate simulated images of the EB wormhole using GRRT simulations, taking into account realistic radiation processes such as synchrotron emission, and compare the resulting image structure with current EHT observations.

This paper is organized as follows.
In Sec.~\ref{sec:EBWHproperty}, we briefly review the EB wormhole solution and the null geodesics in this spacetime.
In Sec.~\ref{sec:accretion}, we derive the spherically symmetric steady-state accretion flow that illuminates the central object. 
In Sec.~\ref{sec:method}, we show the basic equations of the GRRT simulations and the numerical methodology to obtain the simulated image.
In Sec.~\ref{sec:results}, we discuss the simulated images of the EB wormhole and the Schwarzschild black hole, and their differences.
We also compare the resulting image structures with the current EHT results. 
Section~\ref{sec:summary} is devoted to a summary and discussion.
We use the units $c=G=h=1$.

\section{Properties of Ellis-Bronnikov wormhole}
\label{sec:EBWHproperty}

\subsection{Theory and metric}
We consider a massless phantom scalar field $\Phi$ minimally coupled to GR. 
The action is given by 
\begin{align}
    S = 
    \frac{1}{16 \pi}
    \int \rmd^4 x
    \sqrt{-g}
    \brb{ R + 8\pi (\nabla_\mu \Phi)(\nabla^\mu \Phi)}.
\end{align}
The field equations can be written as follows:
\begin{align}
    &R_{\mu \nu} = - 8 \pi (\nabla_{\mu} \Phi) (\nabla_{\nu} \Phi),
    \\
    &\nabla_\mu \nabla^\mu \Phi = 0.
\end{align}

In this system, there is a class of static and spherically symmetric spacetimes known as the Ellis-Bronnikov (EB) wormhole. 
The phantom scalar field and the line element of the EB wormhole are given by 
\begin{align}
    \Phi 
    &=
    \sqrt{\frac{M^2 + \ell^2}{4 \pi \ell^2}}
    \brb{\tan^{-1} \bra{\frac{R}{\ell}} - \frac{\pi}{2}},
    \\
    \rmd s^2
    &=
    - e^{f} \rmd t^2
    + e^{-f}
    \brb{
    \rmd R^2 + 
    (R^2 + \ell^2)
    (\rmd \theta ^2 + \sin^2 \theta \rmd \phi^2)
    },
    \label{eq:metric}
\end{align}
with 
\begin{align}
    f(R) 
    = 
    \frac{2M}{\ell}
    \brb{\tan^{-1} \bra{\frac{R}{\ell}} - \frac{\pi}{2}}, \label{eq:fR}
\end{align}
where $\ell$ is a constant, and $M$ is the ADM mass in the asymptotic flat region for $R \to + \infty$. 
We note that for $R \to - \infty$, the ADM mass of the EB wormhole is given by $-M e^{\pi M / \ell}$. 
This implies that the ADM mass associated with the $R \to - \infty$ asymptotic flat region is negative if $M$ is positive. 
In the EB wormhole, there are two asymptotic regions for $R \to \pm \infty$. 
These asymptotic regions are connected by a throat located at $R_{\rm th} = M$. 
For $R \to + \infty$, since the function $f$ goes to zero, the metric \eqref{eq:metric} reduces to the Minkowski metric. 
On the other hand, for $R \to - \infty$, the metric coincides with the Minkowski metric after the following coordinate transformations:
\begin{align}
    \tilde{t} \coloneqq e^{-\pi M / \ell} t,
    \qquad
    \tilde{R} \coloneqq e^{\pi M / \ell} R, 
    \qquad
    \tilde{\ell} \coloneqq e^{\pi M / \ell} \ell.
\end{align}

For $M = 0$, the EB wormhole describes a symmetric wormhole, which is called the Ellis wormhole. 
The metric of the Ellis wormhole is given by
\begin{align}
    \rmd s^2
    =
    - \rmd t^2 
    + \rmd R^2
    + (R^2 + \ell^2) (\rmd \theta^2 + \sin^2 \theta \rmd \phi^2). 
\end{align}
In the Ellis wormhole, a throat is located at $R = 0$. 
Images of the Ellis wormhole illuminated by steady-state spherical dust fluids have been reported in \cite{Ohgami:2015nra}. 
In this paper, we extend this analysis to the EB wormhole ($M \neq 0$), considering illumination by steady-state spherical polytropic fluids. 

\subsection{\texorpdfstring{Geodesics and the photon sphere}{Geodesics and the photon sphere}}

Here, we analyze geodesics of freely falling test particles in the EB wormhole. 
In this subsection, we consider particles moving in the equatorial plane, i.e., $\theta = \pi / 2$. 

The Lagrangian of a test particle is defined by
\begin{align}
    2\mathcal{L}
    =
    g_{\mu \nu} \dot{x}^{\mu} \dot{x}^{\nu} = - \epsilon, \label{eq:lagrangian}
\end{align}
where the dot denotes derivatives with respect to an affine parameter, and $\epsilon = 1$ for a timelike particle, while $\epsilon=0$ for a massless particle. 
Due to the symmetries of the spacetime, there are two Killing vectors $(\partial_{t})^{\mu}$ and $(\partial_{\phi})^{\mu}$. 
As a result, the energy $E$ and the angular momentum $L$ of a particle are conserved, and the constants of motion are given by
\begin{align}
    E = e^{f} \dot{t},
    \qquad
    L = e^{-f}(R^2 + \ell^2) \dot{\phi}.
\end{align}
From the normalization of the four velocity, we introduce the effective potential as follows:
\begin{align}
    \dot{R}^2 + V_{\rm eff}(R)
    &= 
    E^2, 
    \\
    V_{\rm eff}(R) 
    &\coloneqq 
    e^{f} 
    \bra{
    \epsilon + e^{f} \frac{L^2}{R^2 + \ell^2}
    }.
\end{align}

We analyze circular orbits for a particle. 
A particle in a circular orbit satisfies $\dot{R}= 0$ and $\ddot{R}=0$. 
The condition $\dot{R}=0$ leads to 
\begin{align}
    V_{\rm eff} = E^2. 
    \label{eq:circcon1}
\end{align}
On the other hand, the condition $\ddot{R}=0$ leads to 
\begin{align}
    V_{\rm eff}^{\prime} = 0,
    \label{eq:circcon2}
\end{align}
where the prime denotes the derivative with respect to $R$. 
Therefore, the locations of the circular orbits correspond to those of the stationary points of $V_{\rm eff}$.

Since the images calculated in this paper are formed by photons, we hereafter restrict our attention to massless particles ($\epsilon = 0$).
The conditions for a circular orbit of a massless particle are $\dot{R}=\ddot{R}=0$, i.e., Eqs.~\eqref{eq:circcon1} and \eqref{eq:circcon2}. 
The effective potential of a massless particle is 
\begin{align}
    V_{\rm eff} = e^{2f} \frac{b^2}{R^2 + \ell^2},
\end{align}
where $b \coloneqq L/E$ is the impact parameter. 
We note that a circular orbit of a massless particle is called the photon sphere. 
Since the radius of the photon sphere corresponds to the local maximum of $V_{\rm eff}$, we obtain the radius of the photon sphere from $V_{\rm eff}^{\prime} = 0$ as follows:
\begin{align}
    R_{\rm ps} \coloneqq 2M. 
\end{align}
In addition, by imposing the conditions $\dot{R} = \ddot{R} = 0$, the critical impact parameter for a massless particle that asymptotically approaches the photon sphere is given by 
\begin{align}
    b_{\rm ps} \coloneqq e^{\frac{-2M}{\ell} \brb{\tan^{-1}(2M/\ell)-\pi/2}} \sqrt{4M^2 + \ell^2}. 
\end{align}

For a fixed mass $M$, the critical impact parameter $b_{\rm ps}$ is a monotonically increasing function of $\ell$. 
In the limit $\ell \to 0$,
\begin{align}
    \lim_{\ell \to 0} b_{\rm ps} = 2 e M \simeq 5.44 M.
    \label{eq:bpslimit}
\end{align}
Since this value exceeds the Schwarzschild value $b_{\rm ps}^{\rm Sch} = 3\sqrt{3} M \simeq 5.20 M$, the inequality $b_{\rm ps} > b_{\rm ps}^{\rm Sch}$ holds for $\ell \ge 0$. 
In other words, for a given mass $M$, the size of the bright ring structure of the image for the EB wormhole is always larger than that for the Schwarzschild BH. 
Therefore, in order to match the size of the bright ring structure for the EB wormhole with that for the Schwarzschild BH, we should consider the EB wormhole with a smaller mass. 
We will consider this situation in Sec.~\ref{sec:method}.

\section{\texorpdfstring{Spherically symmetric steady-state accretion flows}{Spherically symmetric steady-state accretion flows}}
\label{sec:accretion}

\subsection{Thermodynamics}

We assume the flow to be adiabatic and free of dissipative processes.
The entropy per particle $\sigma$ is therefore conserved along the flow: $u^{\mu} \nabla_{\mu} \sigma = 0$, where $u^{\mu}$ is the four-velocity of the fluid. 
For an adiabatic fluid, the first law of thermodynamics can be written as
\begin{align}
    \rmd \bra{\frac{\rho}{n}} = - p\, \rmd \bra{\frac{1}{n}},
    \label{eq:therm1stlaw}
\end{align}
where $\rho$, $p$, $n$ are the total energy density, the pressure, and the number density measured in the fluid rest frame, respectively. 
Eq. \eqref{eq:therm1stlaw} leads to the following relations 
\begin{align}
    \rmd p = n \rmd h,
    \qquad
    \rmd \rho = h \rmd n,
    \label{eq:dpdrho}
\end{align}
where $h$ is the enthalpy per particle defined by
\begin{align}
    h \coloneqq \frac{\rho + p}{n}. 
    \label{eq:defenthalpy}
\end{align}
For the later discussion, we introduce the sound speed of the fluid. 
Since we consider the adiabatic flow, the sound speed can be expressed as 
\begin{align}
    a^2 \coloneqq \frac{\rmd p}{\rmd \rho} = \frac{\rmd \ln h}{\rmd \ln n}, 
\end{align}
where we used Eq. \eqref{eq:dpdrho}.

\subsection{Steady-state accretion flows in a static and spherically symmetric spacetime}

We consider a static and spherically symmetric spacetime described by the following metric:
\begin{align}
    \rmd s^2 = 
    - A(R) \rmd t^2 
    + B(R) \rmd R^2
    + C(R) (\rmd \theta^2 + \sin^2 \theta \rmd \phi^2). 
\end{align}
In addition, we consider a perfect fluid whose energy-momentum tensor is given by
\begin{align}
    T^{\mu \nu} = \bra{\rho + p} u^{\mu} u^{\nu} + p g^{\mu \nu}.
\end{align}
Since we focus on a spherically symmetric steady-state accretion flow, the four-velocity of the fluid satisfies $u^{\mu} = (u^t, u^{R}, 0, 0)$. 

Here, we introduce an orthonormal tetrad $\{e^{\mu}{}_{\hat{\alpha}} \}$ attached to a static observer. 
The tetrad is given by
\begin{align}
    e^{\mu}{}_{\hat{\alpha}}
    = 
    {\rm diag} \bra{\frac{1}{\sqrt{A}}, \frac{1}{\sqrt{B}}, \frac{1}{\sqrt{C}}, \frac{1}{\sqrt{C} \sin \theta}}.
\end{align}
This tetrad satisfies $\eta_{\hat{\alpha} \hat{\beta}} = g_{\mu \nu} e^{\mu}{}_{\hat{\alpha}} e^{\mu}{}_{\hat{\beta}}$, where $\eta_{\hat{\alpha} \hat{\beta}}$ is the Minkowski metric. 
The four-velocity measured by a static observer is obtained by projection onto the tetrad as $u^{\hat{\alpha}} = e^{\hat{\alpha}}{}_{\mu}u^{\mu}$. 
In the local frame of a static observer, the four-velocity takes the special relativistic form 
\begin{align}
    u^{\hat{\alpha}} = \bra{\gamma, \gamma v, 0, 0},
\end{align}
where $\gamma \coloneqq \bra{1-v^2}^{-1/2}$ is the Lorentz factor and 
\begin{align}
    v \coloneqq \frac{u^{\hat{R}}}{u^{\hat{t}}} = \sqrt{\frac{B}{A}} \frac{u^R}{u^t} 
\end{align}
is the radial velocity of the fluid measured by a static observer.

Assuming that the number density of particles is conserved, we can find a conserved quantity from the continuity equation $\nabla_{\mu}(n u^{\mu}) = 0$ as 
\begin{align}
    \dot{N} 
    \coloneqq
    - 4\pi \sqrt{A}C n \gamma v,  
    \label{eq:conservedNdot}
\end{align}
where we define the sign of $\dot{N}$ such that $\dot{N}$ is positive for the accretion flow in which $v < 0$. 
Since we consider the static and spherically symmetric spacetime, there is another conserved quantity associated with the timelike Killing vector $(\partial_{t})^{\mu}$. 
With the timelike Killing vector $(\partial_{t})^{\mu}$, the Noether current is given by $J^{\mu} \coloneqq - T^{\mu}{}_{\nu} (\partial_{t})^{\nu}$. 
From the conservation equation for the Noether current $\nabla^{\mu} J_{\mu} = 0$, we can find the conserved quantity as
\begin{align}
    \mathcal{E}^{2} 
    \coloneqq
    h^2 \gamma^2 A.
    \label{eq:conservedEdot}
\end{align}
This quantity represents the relativistic generalization
of the Bernoulli constant and reduces to the classical
Bernoulli constant in the Newtonian limit.

The solutions of the flow are described as the curves in the phase space $(R,v)$. 
These solutions can be obtained by integrating the following equations: 
\begin{align}
     \frac{\rmd}{\rmd \tau} 
     \begin{pmatrix}
         R \\ v
     \end{pmatrix}
         =
         \begin{pmatrix}
             \partial_{v} \\ -\partial_{R}
         \end{pmatrix}
         \mathcal{E}^2(R,v),
         \label{eq:drdtdvdt}
\end{align}
where $\tau$ is a parameter characterizing the curves. 
At the critical point, the right-hand side of Eq.~\eqref{eq:drdtdvdt} vanishes. 
This leads to 
\begin{align}
    \frac{2 A h^2 }{v\bra{1-v^2}^2} \bra{v^2 - a^2} = 0,
    \label{eq:calN}
    \\
    \frac{2 A h^2}{1-v^2} \bra{ \frac{1-a^2}{2 } \frac{A^{\prime}}{A} - a^2 \frac{C^{\prime}}{C}} = 0.
    \label{eq:calD}
\end{align}
From Eq.~\eqref{eq:calN}, it follows that $v^2 = a^2$ at the critical point. 
In other words, the critical point coincides with the sonic point. 
In what follows, we use the notation $(R_{\rm c}, v_{\rm c})$ to denote the critical point. 
We can classify the critical point into two types, i.e., a saddle type or an extremum type. 
In this study, we consider an accretion flow that vanishes at the $R \to +\infty$, transitions from subsonic to supersonic at the sonic point, and then flows into the $R < 0$ region. 
Such accretion flows can be realized when the critical points are saddle type.

\subsection{Application to polytropic fluid}
\label{subsec:polytropic fluid}

In the present paper, we consider polytropic fluids as an explicit example. 
The equation of state (EOS) for polytropic fluids is given by
\begin{align}
    p = K n^{\Gamma},
\end{align}
where $K$ is a constant that is determined by the thermodynamic properties of the fluid, while $\Gamma$ is the adiabatic index. 
In order to calculate a shadow image, we need to find the temperature, the number density, and the radial velocity. 
For the adiabatic fluids, the energy density can be written as
\begin{align}
    \rho = m n + \frac{p}{\Gamma -1}, 
    \label{eq:enegydensityinn}
\end{align}
where $m$ is a mass of an accreting particle. 
Substituting Eq.~\eqref{eq:enegydensityinn} into Eq.~\eqref{eq:defenthalpy}, we find the pressure as follows:
\begin{align}
    p = \frac{\Gamma - 1}{\Gamma} n \bra{h - m}.
    \label{eq:pressureinn}
\end{align}
By using the EOS $p=n k_{\rm B} T$, we finally obtain the expression for the temperature
\begin{align}
    T = \frac{\Gamma -1}{k_{\rm B} \Gamma} \bra{\frac{\mathcal{E}}{\gamma \sqrt{A}} - m}, 
    \label{eq:temperaturewithv}
\end{align}
where we used Eq.~\eqref{eq:conservedEdot}. 
The number density can be calculated from Eq.~\eqref{eq:conservedNdot}: 
\begin{align}
    n = - \frac{\dot{N}}{4 \pi \sqrt{A}C\gamma v}. 
    \label{eq:numberdensitywithv}
\end{align}
Therefore, we can calculate the temperature and the number density if we find the radial velocity $v$. 
To find the radial velocity, we need to solve the following differential equation
\begin{align}
    \frac{\rmd v}{\rmd R} = 
    \frac{- \partial_{R} \mathcal{E}^2}{\partial_v \mathcal{E}^2}. 
    \label{eq:dvdR}
\end{align}
To construct the global transonic solution, we first determine the radial velocity at the sonic point. 
The slope at the sonic point is obtained from a linear analysis of the flow equations. 
Using these values, we slightly shift the radial position to both sides of the sonic point and integrate the equations inward and outward to obtain the full radial velocity profile.

To classify the accretion flows, we linearize the dynamical system around the critical point by expanding the solution as $(R, v) = (R_{\rm c} + \delta R, v_{\rm c} + \delta v)$. 
Expanding Eq.~\eqref{eq:drdtdvdt} around the critical point, the linearized equation is given by
\begin{align}
    \frac{\rmd}{\rmd \tau}
    \begin{pmatrix}
        \delta R \\ \delta v
    \end{pmatrix}
    =
    \mathcal{J} 
    \begin{pmatrix}
        \delta R \\ \delta v
    \end{pmatrix},
\end{align}
where we introduced the linearization matrix $\mathcal{J}$ defined by
\begin{align}
    \mathcal{J}
    \coloneqq
    \begin{pmatrix}
        \partial_{R} \partial_{v} & \partial_{v}^{2} \\
        - \partial_{R}^{2} & -\partial_{R} \partial_{v} 
    \end{pmatrix}
    \mathcal{E}^{2} (R_{\rm c}, v_{\rm c}).
\end{align}
The nature of the critical point is determined by the eigenvalues of the linearization matrix.
If the eigenvalues are real and have opposite signs, the critical point is of saddle type. 
Therefore, $\det \mathcal{J} < 0$ is required for realizing transonic accretion flows. 
In this case, there exist two distinct directions in phase space along which the flow can pass through the critical point.
By contrast, if the eigenvalues have the same sign or form a complex conjugate pair, the flow cannot represent a physically viable transonic accretion flows passing through the critical point.

\begin{figure}
  \begin{tabular}{cc}
    \begin{minipage}{\hsize}
      \begin{center}
        \includegraphics[width=\textwidth]{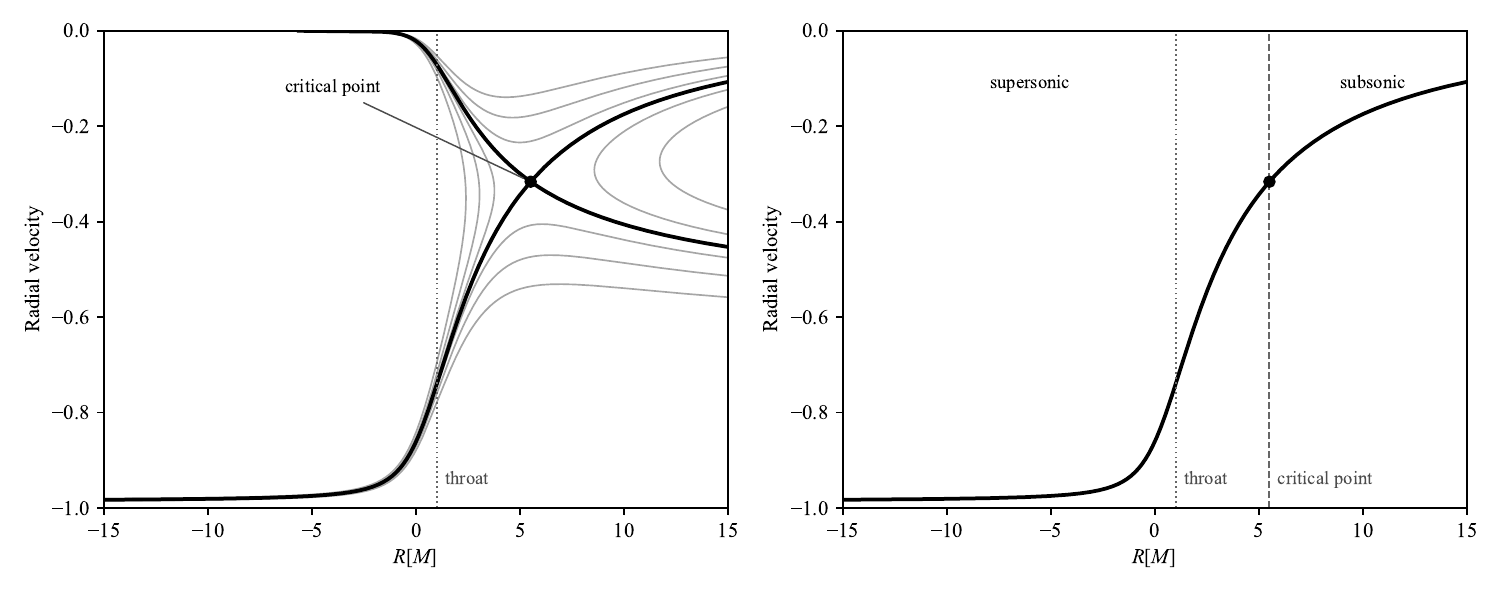}
         \caption{Left: level curves of $\mathcal{E}^2(R,v)$ in the phase space of the EB wormhole for $\ell = 2M$ and $\Gamma = 4/3$, with the sound speed at the critical point fixed to $v_{\rm c}^2 = 1/10$. The thick curve is the level curve passing through the critical point, and the black dot denotes the critical point, which is located at $R_{\rm c} = 5.5M$ and is of saddle type. 
         The thin curves are level curves for neighboring values of $\mathcal{E}^2$. Right: the transonic accretion flow adopted in this paper, i.e., the branch of the thick curve that is subsonic at large $R$ and supersonic inside the critical point. The dotted and dashed vertical lines indicate the locations of the throat and the critical point, respectively.}
         \label{fig:flowphase}
       \end{center}
     \end{minipage}
   \end{tabular}
\end{figure}
Figure~\ref{fig:flowphase} shows the phase space of the EB wormhole for $\ell = 2M$ and $\Gamma = 4/3$.
Since Eq.~\eqref{eq:drdtdvdt} takes a Hamiltonian form, $\mathcal{E}^2(R,v)$ is conserved along each solution, and hence the solutions of the flow are the level curves of $\mathcal{E}^2(R,v)$ at a fixed $\dot{N}$. 
As seen in the left panel, the level curves passing through the critical point cross each other. 
This behavior is the characteristic structure of a saddle-type critical point. 
The other level curves either turn around before reaching the throat or never extend to the asymptotic region, and therefore they cannot describe a flow that connects the asymptotic region to the region beyond the throat.
The transonic accretion flow adopted in this paper is the branch shown in the right panel: the flow is subsonic at large $R$, passes through the critical point, and becomes supersonic, falling through the throat into the $R<0$ region.
We emphasize that the radial velocity determines the flow completely, since the temperature and the number density can be constructed from the radial velocity (see Eqs.~\eqref{eq:temperaturewithv} and \eqref{eq:numberdensitywithv}). 
The distributions used in the following sections are obtained in this way, and are not prescribed by hand.

\section{Method} 
\label{sec:method}

\subsection{General relativistic radiative transfer equation}

The GRRT equation is described as 
\begin{align} 
    \frac{\rmd \mathcal{I}}{\rmd \lambda} = \mathcal{J_{\rm e}} - \mathcal{AI}, \label{eq:grrteq}
\end{align}
where $\lambda\ \rm[s\ g^{-1}], \mathcal{I}\ \rm[erg^{-2}cm^{-2}s^{-1}g^{-1}sr^{-1}], \mathcal{J_{\rm e}}\ \rm[erg^{-2}cm^{-2}s^{-2}sr^{-1}]$ and $\mathcal{A}\ \rm[s^{-1}g]$ are an affine parameter, the invariant intensity (invariant distribution function of photons in a 6-dimensional phase space), invariant emissivity, and invariant absorption coefficient, respectively \citep{Fuerst:2004ii,Younsi:2012ha,Takahashi2017,Takahashi2022}. 
A number of photons $\rmd N$ is given by $\rmd N = \mathcal{I} \rmd V_{\rm (6)}$, where $\rmd V_{(6)}$ is an invariant volume element in a 6-dimensional phase space along a null geodesic shown by Liouville's theorem~\cite{Lindquist:1966igj}. The invariant emissivity and invariant absorption coefficient show creation and annihilation of photons in a 6-dimensional phase space volume element. 
Here, $\rmd \lambda$ is related to the line element $\rmd s$ as $\rmd s = \nu \rmd \lambda$, where $\nu$ is a frequency measured in a local Minkowski frame.
The invariant variables are defined by 
\begin{align} 
    &\mathcal{I} \coloneqq \frac{I_\nu}{{\nu}^3}, \\
    &\mathcal{J_{\rm e}} \coloneqq \frac{j_{\nu}}{{\nu}^2}, \\
    &\mathcal{A} \coloneqq \nu \kappa_{\nu},
\end{align}
where $I_{\nu}\ \rm[erg\ cm^{-2}s^{-1}sr^{-1}Hz^{-1}]$, $j_{\nu}\ \rm[erg\ cm^{-3}s^{-1}sr^{-1}Hz^{-1}]$, and $\kappa_{\nu}\ \rm[cm^{-1}]$ are the intensity, the emissivity, and the absorption coefficient measured in a local Minkowski frame, respectively. 
By substituting $\mathcal{I}=I_{\nu_{\rm{obs}}}/\nu_{\rm obs}^3,\ \rmd s=\nu_{\rm obs} \rmd \lambda$ on the left-hand side and $\mathcal{I}=I_{\nu_{\rm{rest}}}/\nu_{\rm rest}^3,\ \mathcal{J_{\rm e}}=j_{\nu_{\rm{rest}}}/\nu_{\rm rest}^2,\ \mathcal{A}=\nu_{\rm rest} \kappa_{\nu_{\rm rest}}$ on the right-hand side of Eq. (\ref{eq:grrteq}), the GRRT equation can be written as
\begin{align}
    \frac{\rmd I_{\nu_{\rm{obs}}}}{\rmd s} = g^3 \left( j_{\nu_{\rm{rest}}} - \kappa_{\nu_{\rm{rest}}} I_{\nu_{\rm{rest}}} \right), \label{eq:grrt2}
\end{align}
where $I_{\nu_{\rm{obs}}}$ is the intensity measured at the frequency $\nu_{\rm obs}$ in the observer's frame, while $I_{\nu_{\rm{rest}}}$, $j_{\nu_{\rm{rest}}}$, and $\kappa_{\nu_{\rm{rest}}}$ are the intensity, the emissivity, and the absorption coefficient in the fluid rest frame, respectively. 
Note that $I_{\nu_{\rm obs}}$ is defined along the geodesic, and its value at the screen is used to evaluate the brightness temperature. 
Here, $g \coloneqq \nu_{\rm obs}/\nu_{\rm rest}$ is the redshift factor, which represents the frequency shift including both the gravitational redshift and the Doppler shift by the fluid motion, where $\nu_{\rm obs}$ and $\nu_{\rm rest}$ are the photon frequencies in the observer's frame and the fluid rest frame, respectively. Furthermore, we ignore the synchrotron self-absorption (i.e., $\kappa_{\nu_{\rm{rest}}} = 0$) because we assume that the accretion flows are sufficiently optically thin.

In this work, we assume that the accreting matter emits synchrotron radiation. 
The angle-averaged thermal synchrotron emissivity in the fluid rest frame is 
approximated by \citep{Mahadevan1996}
\begin{align} 
    j_{\nu_{\rm rest}} = \frac{n_{\rm e} e^2 \nu_{\rm rest}}{\sqrt{3} K_2(\Theta_{\rm e}^{-1})} \mathcal{M}(w_{\rm cs}), \label{eq:jnusynch}
\end{align}
where $n_{\rm e}$ is the electron number density, $e$ is the electric charge, and $K_2$ is the modified Bessel function of the second kind. 
The dimensionless electron temperature is defined as 
$\Theta_{\rm e} \coloneqq k_{\rm B} T_{\rm e}/m_{\rm e}$, with $T_{\rm e}$ and 
$m_{\rm e}$ being the electron temperature and mass, respectively. 
The function $\mathcal{M}(w_{\rm cs})$ is given by
\begin{align}
    \mathcal{M}(w_{\rm cs}) &= \frac{4.0505}{w_{\rm cs}^{1/6}} \left( 1 + \frac{0.4}{w_{\rm cs}^{1/4}} + \frac{0.5316}{w_{\rm cs}^{1/2}} \right) \exp (-1.8899 w_{\rm cs}^{1/3}),  \label{eq:M_in_jnu}
\end{align}
where $w_{\rm cs} \coloneqq \nu_{\rm rest}/\nu_{\rm cs}$ is the frequency normalized by the characteristic synchrotron frequency $\nu_{\rm cs} \coloneqq 3eB_{\rm mag}\Theta_{\rm e}^2/4\pi m_{\rm e}$, and $B_{\rm mag}$ is the magnetic field strength measured in the fluid rest frame. 
In general, at the EHT observing frequency of 230 GHz, 
the horizon-scale emission of M87* is commonly modeled as the synchrotron emission from a relativistic thermal electron distribution \citep{EHT5, Mizuno2021, Zhang2024}. 
The thermal synchrotron emissivity represented by Eqs. (\ref{eq:jnusynch}) and (\ref{eq:M_in_jnu}) is a standard formula to calculate images of optically thin hot accretion flow (radiatively inefficient accretion flow) around compact objects, and is widely used in various theoretical studies \citep{Pu2016_code, Pandya2016, Dexter2016, EventHorizonTelescope:2020tst, Kawashima2023, Takahashi2026}. 

The magnetic field strength is derived by
\begin{align}
    B_{\rm mag}=\sqrt{\frac{8\pi n_{\rm e} k_{\rm B} T_{\rm e}}{\beta_{\rm p}}}, \label{eq:Bmag}
\end{align}
where $\beta_{\rm p}$ is the ratio of the electron pressure to the magnetic pressure. 
We set $\beta_{\rm p}=0.1$ throughout this paper. 
Global GRMHD simulations show that the ratio of the total gas pressure to the magnetic pressure is approximately an order of unity in the strongly magnetized hot accretion flow \citep{Ressler2015, Porth2019, Chael2019, Anantua2024}. 
When the ratio of the total gas pressure to the magnetic pressure is $\mathcal{O}(1)$, $\beta_{\rm p}$ is $\mathcal{O}(0.1)$ because the electron temperature is smaller than the proton temperature. Thus, $\beta_{\rm p}=0.1$ used in this paper is physically reasonable. In addition, the magnetic field strength around the compact object by setting $\beta_{\rm p} = 0.1$ is to be consistent with those estimated in the EHT observations ($B_{\rm mag}\simeq 1-30\ \rm{G}$).

\subsection{Numerical setup} 

We simulate the shadow images of the Schwarzschild BH and the EB wormhole.
For the Schwarzschild BH, we adopt a mass $M^{\mathrm{fid}} \coloneqq 6.20\times 10^9 \ M_\odot$ ($M_\odot$: the solar mass), which we refer to as the fiducial mass. 
The value of $M^{\rm fid}$ is the mass of M87$^*$ adopted in the EHT analysis~\cite{EHT1, EHT6}. 
We also calculate the shadow image of the EB wormhole with the fiducial mass,
and find that its shadow size is larger than that of the Schwarzschild BH.
To match the shadow size of the EB wormhole with that of the Schwarzschild BH,
we also consider the EB wormhole with $M^{\mathrm{low}} \coloneqq 5.27\times 10^9 \ M_\odot$. 
Hereafter, we refer to the EB wormholes with $M^{\mathrm{fid}}$ and $M^{\mathrm{low}}$ as the fiducial-mass EB wormhole and the low-mass EB wormhole, respectively. 
The value of $M^{\mathrm{low}}$ is determined numerically: we compute images for several masses and adopt the one for which the size of the bright ring-like structure coincides with that of the Schwarzschild BH with $M^{\mathrm{fid}}$. 
In addition, we set $\ell=2M$ in Eq.~\eqref{eq:metric}. 
For all cases, the distance $D$ to the object is set to $D=16.8\ {\rm Mpc}$, 
motivated by the EHT observation for M87$^*$ \citep{EHT1}.

\begin{figure}[t]
  \begin{tabular}{cc}
    \begin{minipage}{\hsize}
      \begin{center}
        \includegraphics[width=\linewidth]{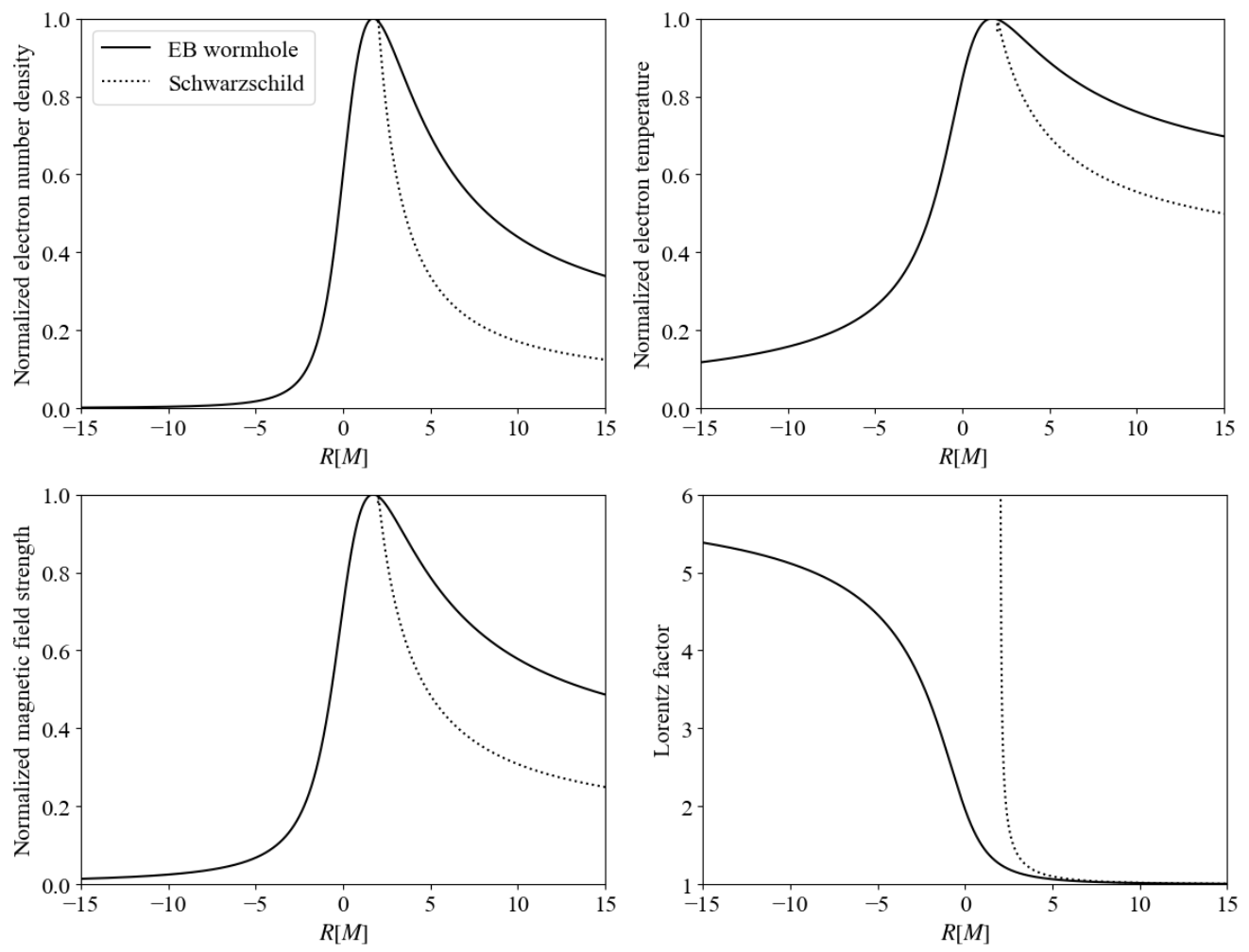}
         \caption{Distributions of the electron number density (top left), the electron temperature (top right), the magnetic field strength (bottom left), and the Lorentz factor (bottom right). 
         The solid and dotted curves correspond to the EB wormhole and the Schwarzschild BH cases, respectively. 
         The electron number density, the electron temperature, and the magnetic field strength are normalized by their respective maximum values. 
         The distributions are shown in
        the ranges used in the GRRT calculation, $-15M < R < 15M$ for the EB wormhole
        and $2M + \delta < R < 15M$ with $\delta = M/200$ for the Schwarzschild BH,
        to which the transonic solutions are truncated.}
         \label{fig:fluid}
       \end{center}
     \end{minipage}
   \end{tabular}
\end{figure}
Following Sec.~\ref{sec:accretion}, we obtain a solution of a spherically symmetric steady-state transonic accretion flow onto the Schwarzschild BH and the EB wormhole. 
We assume polytropic fluids with $\Gamma = 4/3$, where $\Gamma$ is the adiabatic index.  We obtain the transonic solution numerically in the range
$-200M < R < 200M$ for the EB wormhole and $2M + \delta < R < 200M$ with
$\delta = M/200$ for the Schwarzschild BH, where $\delta$ is a small offset
introduced to avoid the numerical divergence at the event horizon.
To obtain the transonic solution, we impose the transonic condition at the
critical point and integrate Eq.~\eqref{eq:dvdR} inward and outward as mentioned in Sec.~\ref{subsec:polytropic fluid}. 
In the GRRT calculation, we truncate this solution to $-15M < R < 15M$ for
the EB wormhole and $2M + \delta < R < 15M$ for the Schwarzschild BH; that is,
we focus on the emission from the vicinity of the central object, where the
differences between the two spacetimes appear (Sec.~\ref{subsec:images}),
and neglect the emission from the accreting matter outside these ranges. 
For both the Schwarzschild BH and the EB wormhole, the electron number density, the temperature, the magnetic field strength, and the Lorentz factor are shown in Fig.~\ref{fig:fluid}. 
The maximum values of the electron number density and the electron 
temperature are set to $1.7\times 10^5 \ {\rm cm^{-3}}$ and 
$5.0\times 10^{10}\ {\rm K}$ for the fiducial-mass EB wormhole, 
$2.4\times 10^5 \ {\rm cm^{-3}}$ and $5.0\times 10^{10}\ {\rm K}$ 
for the low-mass EB wormhole, and $5.6\times 10^5 \ {\rm cm^{-3}}$ 
and $1.0\times 10^{11}\ {\rm K}$ for the Schwarzschild BH, 
respectively. 
These parameters are chosen so that the total flux of the image is approximately $0.5\ {\rm Jy}$\footnote{``Jy'' is a unit of flux 
density ($1\ {\rm Jy} = 10^{-23}\ {\rm erg \cdot s^{-1} \cdot 
cm^{-2} \cdot Hz^{-1}}$).}, which is consistent with the estimates from the EHT polarization observations 
\citep{EHT7, EHT8}.

To integrate the geodesic equation and generate realistic EB wormhole images, we use the GRRT code: \texttt{CARTOON} \citep{Takahashi2022}. 
The geodesic equation is integrated backward in time along the geodesic from $R=1000M$. 
The interval along the geodesic (i.e., time step to integrate the geodesic equation), $\Delta s=\nu \Delta \lambda$, is set to be $10^{-2}M$. 
The viewing angle of the observer is arbitrary because both the spacetime and the accretion flow are spherically symmetric.

\begin{figure}[t]
  \begin{tabular}{cc}
    \begin{minipage}{\hsize}
      \begin{center}
        \includegraphics[width=\linewidth]{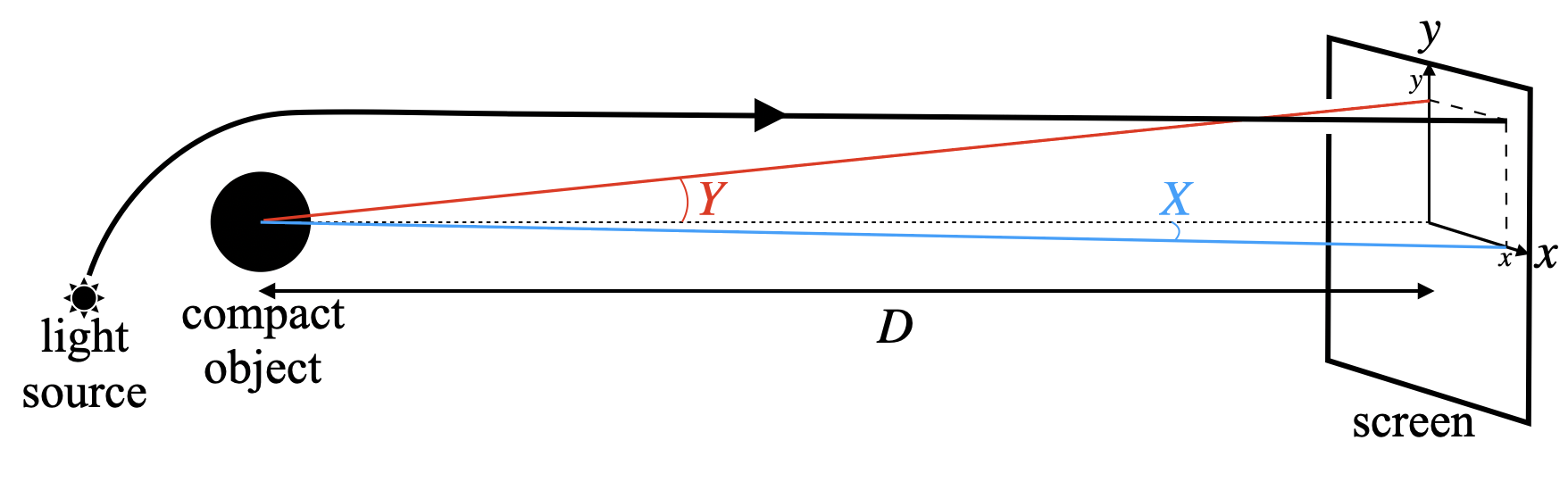}
         \caption{Schematic picture of the definition of the observer's screen. 
         The screen is spanned by the coordinates $(x,y)$. 
         The angular coordinates on the observer's sky are defined by $(X,Y)=(x/D,y/D)$, where $D=16.8$ Mpc is the distance to the compact object. 
         The dotted line corresponds to $X=Y=0$, and the direction toward the center of the compact object.}
         \label{fig:screen}
       \end{center}
     \end{minipage}
   \end{tabular}
\end{figure}
Next, we define an observer's screen (hereafter, just ``screen") on which the observer views a compact object (see also Fig.~\ref{fig:screen}). 
The screen is spanned by the coordinates $(x, y)$. 
Then, let us define the angular coordinates $(X,Y)$ as $(X,Y)=(x/D,y/D)$, where $D = 16.8\,\mathrm{Mpc}$ is the distance to the
compact object\footnote{In the numerical calculation, the null geodesics are traced backward in time from $R = 1000M$, while the screen is assumed to be located at the distance $D$ from the compact object. This treatment is justified as follows. Since the spacetime can be regarded as flat at $R = 1000M$, the geodesics propagate along straight lines parallel to the line of sight beyond this radius, and the intensity is conserved along them because the emission there is neglected. 
Therefore, the intensity distribution on the screen, as a function of $(x, y)$, is the same whether the screen is placed at $R = 1000M$ or at the distance $D$.}.
The screen spans $-10\ M/D \leq X, Y \leq 10\ M/D$, and is uniformly divided into $512 \times 512$ pixels. 
We integrate the geodesic equations backward in time from the center of each pixel on the screen. 
We also assume that the observed frequency is $\nu_{\rm obs} = 230\ {\rm GHz}$ . 
We set this observed frequency because this is the same frequency as EHT observations, and we directly compare our simulated images and EHT's images.

\section{Results}
\label{sec:results}

\subsection{Simulated images}
\label{subsec:images}

\begin{figure}
  \begin{tabular}{cc}
    \begin{minipage}{\hsize}
      \begin{center}
        \includegraphics[width=\linewidth]{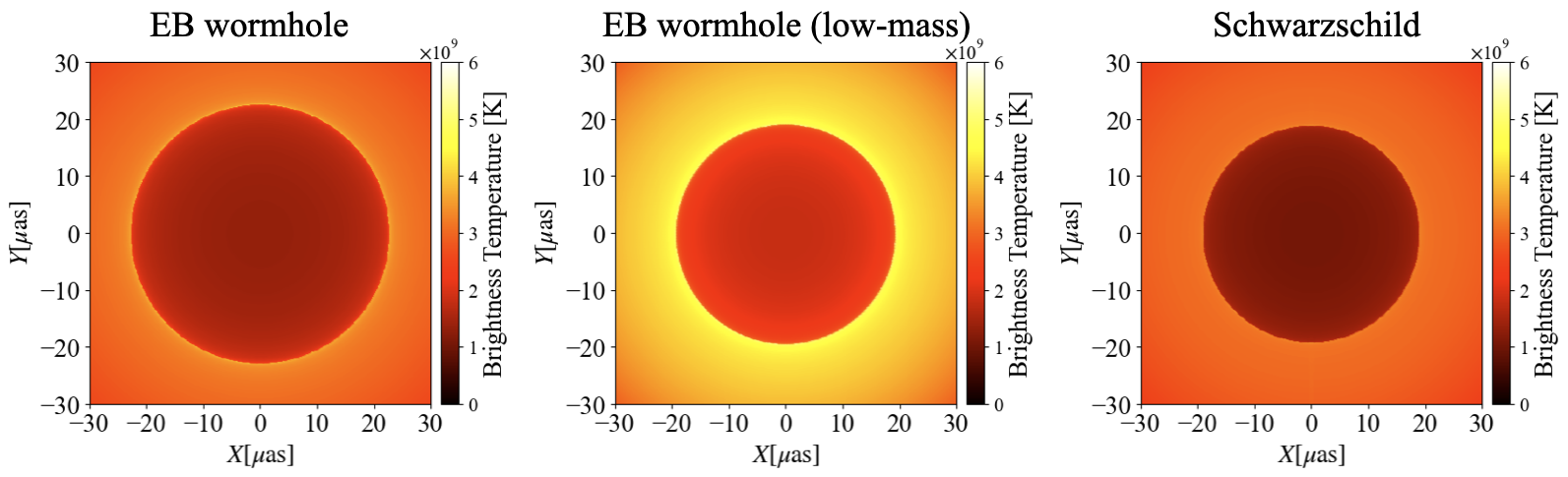}
         \caption{Simulated images of the fiducial-mass EB wormhole (left), the low-mass EB wormhole (center), and the Schwarzschild BH (right), respectively. 
         The color denotes the brightness temperature $T_{\rm b}$ defined by Eq.~\eqref{eq:Tb}.}
         \label{fig:image}
       \end{center}
     \end{minipage}
   \end{tabular}
\end{figure}

Figure~\ref{fig:image} shows the simulated images for the fiducial-mass EB wormhole (left panel), the low-mass EB wormhole (center panel), and the Schwarzschild BH (right panel). 
The color scale represents the brightness temperature $T_{\rm b}$, which is defined by 
\begin{align}
    T_{\rm b} \coloneqq \frac{I_{\nu_{\rm obs}}}{2 \nu_{\rm obs}^2 k_{\rm B}}.
    \label{eq:Tb}
\end{align}
The brightness temperature can be evaluated from the intensity $I_{\nu_{\rm obs}}$ at the screen, which is obtained by solving the GRRT equation. 
All integrations to obtain the null geodesics and the brightness temperature distribution on the screen are conducted with GRRT code: \texttt{CARTOON} \cite{Takahashi2022}. 

All images exhibit a central dark region. 
We call this dark region the shadow region. 
The shadow region is formed by the photons with angular momenta smaller than those of photons on the unstable circular orbit. 
On the other hand, all images also exhibit a bright ring-like structure. We call this bright ring-like structure the photon ring. 
The photon ring is formed by the photons orbiting around the central object multiple times before reaching the observer. 
The diameter of the photon ring for the fiducial-mass EB wormhole is greater than that of the Schwarzschild BH, as discussed in Sec.~\ref{sec:EBWHproperty}.
For the Schwarzschild BH, the radius of the photon ring is $3\sqrt{3} M \sim 5.19M$, while for the fiducial-mass EB wormhole with $\ell=2M$, the radius is about $6.21M$.

\begin{figure}
  \begin{tabular}{cc}
    \begin{minipage}{\hsize}
      \begin{center}
        \includegraphics[width=0.7\linewidth]{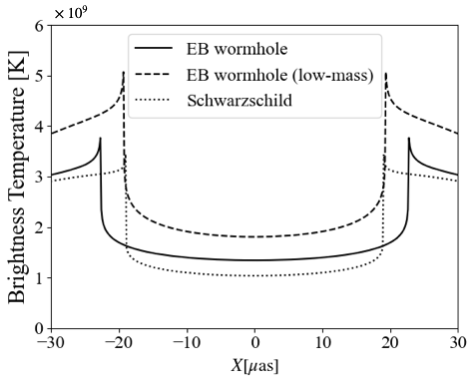}
         \caption{
         The brightness temperature profiles along the $Y=0$ of the simulated images shown in Fig.~\ref{fig:image}.
         The solid, dashed, and dotted curves are the fiducial-mass EB wormhole case, the low-mass EB wormhole case, and the Schwarzschild BH case, respectively. }
         \label{fig:slice}
       \end{center}
     \end{minipage}
   \end{tabular}
\end{figure}
Figure \ref{fig:slice} shows the brightness temperature profiles along the $Y=0$ for the EB wormhole cases (solid and dashed curves) and the Schwarzschild BH case (dotted curve).
There are notable differences between the EB wormhole cases and the Schwarzschild BH case.
The EB wormhole images are brighter than the Schwarzschild BH image, both in the shadow region and at the photon ring.

\begin{figure}
  \begin{tabular}{cc}
    \begin{minipage}{\hsize}
      \begin{center}
        \includegraphics[width=\linewidth]{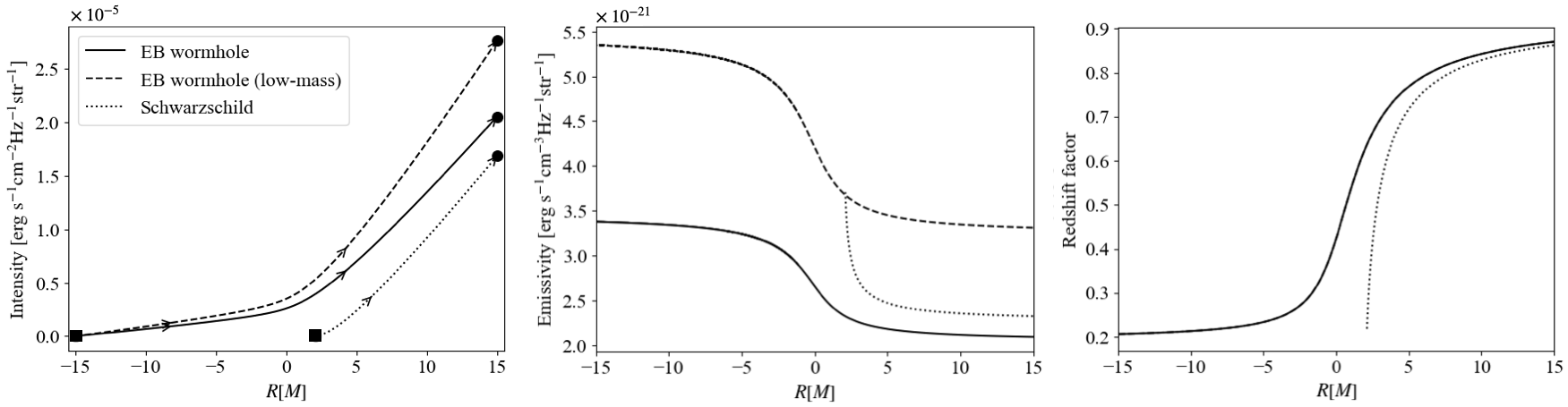}
         \caption{The intensity (left), the emissivity in the fluid rest frame (center), and the redshift factor (right) along the geodesic of a photon that reaches the center of the screen. 
         The solid, dashed, and dotted curves represent the cases of the EB wormhole, the low-mass EB wormhole, and the Schwarzschild BH, respectively. 
         In the left panel, the black square is the point at which the intensity is set to zero. 
         On the other hand, the black circle is the point at which the geodesic reaches $R = 15M$ on its way to the screen; the intensity does not change beyond this point, and hence the value at the black circle corresponds to the intensity at the screen. 
         The arrow indicates the direction of the photon propagation, along which the intensity accumulates.}
         \label{fig:center}
       \end{center}
     \end{minipage}
   \end{tabular}
\end{figure}
The higher brightness temperature of the shadow region in the EB wormhole can be understood as follows.
Figure~\ref{fig:center} shows the intensity $I_{\nu_{\rm obs}}$, the emissivity in the fluid rest frame $j_{\nu_{\rm rest}}$, and the redshift factor $g$ along a null geodesic reaching the center of the screen $(X,Y)=(0,0)$. 
As shown in Eq. (\ref{eq:grrt2}), the intensity at the screen can be obtained by an integration of $g^3 j_{\nu_{\rm rest}} \rmd s$ along the geodesic. 
The integration starts from the black square in Fig.~\ref{fig:center}, where the intensity is set to zero. 
In the left panel of Fig.~\ref{fig:center}, the intensity along the geodesic increases from the black square to the black circle, i.e., along the direction of the photon propagation. 
Since the emission from the region $R>15M$ is neglected, the intensity remains constant beyond the black circle, and hence the intensity at the black circle corresponds to the intensity at the screen. 
The intensity at the screen center for the fiducial-mass EB wormhole is higher than that for the Schwarzschild BH, although the emissivity (shown in the center panel of Fig.~\ref{fig:center}) for the fiducial-mass EB wormhole case is smaller than that for the Schwarzschild BH case.
This is because, in the fiducial-mass EB wormhole case, the accreting matter exists not only in the vicinity of the throat but also beyond it, owing to the absence of the event horizon, and the value of the redshift factor of the EB wormhole spacetime is larger than that of the Schwarzschild BH.

\begin{figure}
  \begin{tabular}{cc}
    \begin{minipage}{\hsize}
      \begin{center}
        \includegraphics[width=\linewidth]{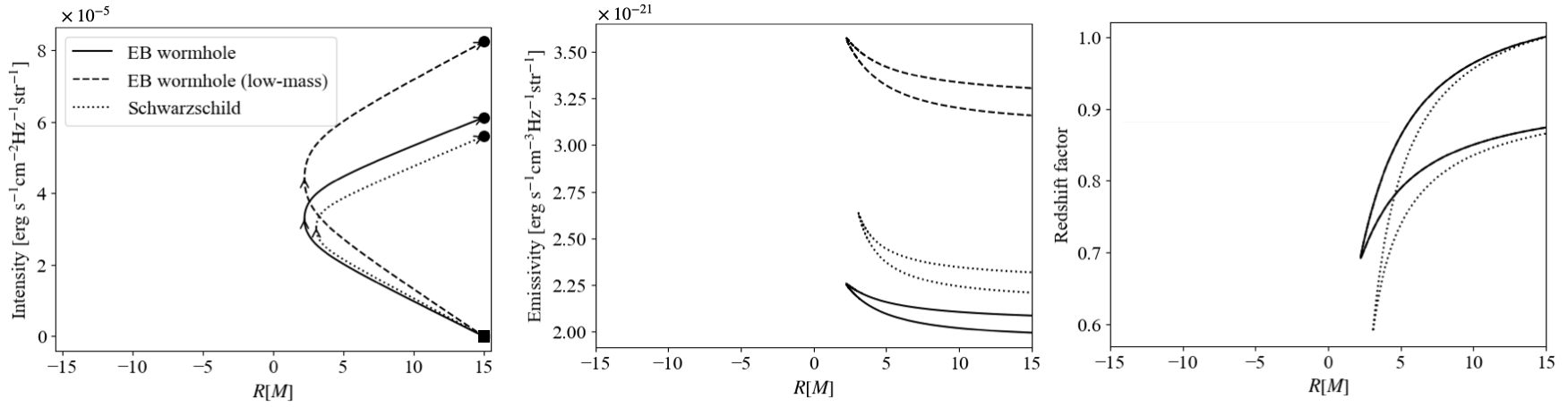}
         \caption{The intensity (left), the emissivity in the fluid rest frame (center), and the redshift factor (right) along the geodesic of a photon that produces the peak intensity. 
         In the left panel, both end points of the geodesic are located at $R = 15M$: the black square is the point at which the intensity is set to be zero, and the black circle is the point at which the geodesic, after orbiting the central object, returns to $R = 15M$ on its way to the screen. 
         The arrow in the left panel indicates the direction of the photon propagation, along which the intensity accumulates; the value at the black circle therefore corresponds to the intensity at the screen.}
         \label{fig:ring}
       \end{center}
     \end{minipage}
   \end{tabular}
\end{figure}

Next, we discuss why the photon ring of the EB wormhole is brighter than that of the Schwarzschild BH. 
To this end, we focus on the null geodesic whose intensity at the screen is maximum. 
Figure~\ref{fig:ring} shows the intensity $I_{\nu_{\rm obs}}$, the emissivity in the fluid rest frame $j_{\nu_{\rm rest}}$, and the redshift factor $g$ along 
such a null geodesic. 
The turning points seen in Fig.~\ref{fig:ring} reflect the trajectories of the geodesics: starting from the black square, approaching and orbiting the central object several times near the photon sphere, and finally propagating outward to reach $R = 15M$ at the black circle and then the screen. 
In the fiducial-mass case, although the emissivity along the geodesic for the Schwarzschild BH is larger than that for the EB wormhole (see the center panel of Fig.~\ref{fig:ring}), the photon ring of the EB wormhole is brighter than that of the Schwarzschild BH, as seen in the left panel of Fig.~\ref{fig:ring}. 
There are two main reasons for this.
The first is that the path length of the photon for the EB wormhole is longer than that for the Schwarzschild BH; the difference in the coordinate time along the geodesic is 
$\Delta t \sim 3.4M$\footnote{The affine parameter length in the 
EB wormhole case is also longer than that in the Schwarzschild BH case by approximately $9.1M$.}. 
The second is the smaller redshift factor in the Schwarzschild BH case compared with the EB wormhole case as shown in the right panel of Fig. \ref{fig:ring}. 
Even a small difference in the redshift factor can result in a significant difference in the intensity, since the intensity scales as $g^3$ in the GRRT equation~\eqref{eq:grrt2}.

For the low-mass EB wormhole, the brightness temperature in both the shadow region and the photon ring becomes larger than that of the Schwarzschild BH. 
This is because the emissivity along the geodesic for the low-mass EB wormhole is larger than that for the Schwarzschild BH, both for the geodesic reaching the screen center (the dashed curve in the center panel of Fig.~\ref{fig:center}) and for the geodesic producing the peak intensity (the dashed curve in the center panel of Fig.~\ref{fig:ring}).

We note that the simulated image of the Ellis wormhole is discussed in Appendix~\ref{app:Ellisimage}. 
For the Ellis wormhole, the region inside the photon ring is brighter than the region outside the photon ring. 
This trend is consistent with the previous work~\cite{Ohgami:2015nra}.

\subsection{Comparison with EHT observations}

\setlength{\tabcolsep}{8pt}
\begin{table}
    \centering
    \begin{tabular}{c|ccc}
         & Total flux [Jy] & Ring diameter [${\rm \mu as}$] & Central depression \\
         \hline
        EHT M87$^*$ & 0.2--1.2 \citep{EHT8} & $42 \pm 3$ \citep{EHT1} & $\miki{> 10}$ \citep{EHT1} \\
        EBWH ($M^{\rm fid}$) & 0.50 & $\simeq 45$ & $\simeq 2.8$ \\
        EBWH ($M^{\rm low}$) & 0.49 & $\simeq 38$ & $\simeq 2.8$ \\
        Schwarzschild BH & 0.49 & $\simeq 38$ & $\simeq 1.9$ \\
    \end{tabular}
    \caption{The total flux, the ring diameter, and the central depression of
    the EHT M87$^*$ result and our simulated images.}
    \label{tab:structure}
\end{table}

Our setup shares several key ingredients with the EHT analysis~\cite{EHT1, EHT5}:
the mass $M^\mathrm{fid}$, the distance $D$, the observing frequency of
$230\,\mathrm{GHz}$, the thermal synchrotron emission, and the total flux of
approximately $0.5\,\mathrm{Jy}$.
However, the EHT analysis is based on GRMHD simulations around the Kerr BH,
whereas our models assume a spherically symmetric transonic accretion flow in
non-rotating spacetimes, neglect the synchrotron self-absorption, and are not
convolved with the beam size of the EHT.

Here, we compare several observables derived from our simulated images with the EHT results. 
Table \ref{tab:structure} shows the total flux, the ring diameter, and the central depression (the ratio of the maximum to the minimum intensity) of our simulated images and the EHT result for $\rm{M87^*}$. 
The total flux and the ring diameter of our simulated images are comparable with the EHT results for M87*. 
However, the central depression is much smaller than the observed lower bound of $10$~\cite{EHT1} for both the EB wormhole and the Schwarzschild BH. 
Thus, our models with the spherically symmetric accretion flow cannot reproduce the EHT image of M87*. 
This means that our models with the spherically symmetric accretion flow are
rejected as a model of M87*, for both the EB wormhole and the Schwarzschild BH.

The small central depression is a consequence of the spherically symmetric
accretion flow adopted in this paper. 
For the axisymmetric fluid distribution, which corresponds to a more realistic situation, it is expected that the intensity of photons whose angular momentum is aligned with that of the accretion flow becomes greater via the Doppler boosting. 
In contrast, the intensity of photons whose angular momentum is in the opposite direction to that of the matter is suppressed by the Doppler deboosting. 
As a result, when there is an axisymmetric fluid distribution, the image structure becomes asymmetric, and the ratio of the maximum to the minimum intensity is enhanced.

\section{Summary and discussion}
\label{sec:summary}

In this paper, we have calculated the shadow images of the EB wormhole and the Schwarzschild BH for spherically symmetric steady-state transonic accretion flows. 
We have taken into account the synchrotron emission and solved the GRRT equation to obtain realistic shadow images. 
For both the EB wormhole and the Schwarzschild BH, the images consist of the central dark region (shadow region) and a bright ring-like structure (photon ring). 
We have found that when the mass is fixed, the shadow region for the EB wormhole is brighter than that for the Schwarzschild BH. 
This is because the photons emitted by the accreting matter that exists not only in the vicinity of the throat but also beyond it can contribute to the intensity in the EB wormhole case. 
In addition, we have found that the photon ring is also brighter for the EB wormhole than for the Schwarzschild BH. 
This is because the path length of the photon that generates the photon ring in the EB wormhole spacetime is longer than that in the Schwarzschild BH spacetime, and the redshift factor in the vicinity of the EB wormhole is larger than that near the Schwarzschild BH. 
In the EB wormhole, the lapse function does not vanish near the object because there is no event horizon, unlike the Schwarzschild BH. Thus, the resulting brightness temperature of the photon ring in the EB wormhole case tends to be larger than that in the Schwarzschild BH case.

We have found that the total flux and the ring diameter are comparable with the EHT results, whereas the central depression is smaller than the observed one for both the EB wormhole and the Schwarzschild BH. Thus, our models with the spherically symmetric accretion flow cannot reproduce the EHT image of M87*, and a non-spherical accretion flow is required for a detailed comparison with the EHT observations.
In order to distinguish the EB wormhole from a BH spacetime, observations with sufficiently high angular resolution are required. 
A promising candidate for such high-resolution observations is a space-VLBI observation, such as the ``Black Hole Explorer" (BHEX) mission \citep{Johnson:2024ttr,Akiyama:2024msp,Kawashima:2024svy} planned for the early 2030s.

Simulations of more realistic images, taking into account more realistic fluid distributions, are left for future work. 
Furthermore, in reality, an axisymmetric accretion flow (accretion disk) forms around the compact object. 
When an accretion disk exists around the compact object, an asymmetric intensity distribution is expected to be realized. Furthermore, in the future, it is also possible to carry out calculations that take into account a more realistic structure of the accretion flow using GRMHD simulations.

\section*{Acknowledgment}

We thank the anonymous referee for valuable comments and suggestions that helped us improve the manuscript. 
Our numerical simulations were conducted with HPE Cray XD2000 at the Center for Computational Astrophysics, National Astronomical Observatory of Japan. This work was supported by Takahashi Industrial and Economic Research Foundation and JSPS KAKENHI Grant Numbers 25K01045, 26K17201 (MMT).


%

\vspace{0.2cm}


\let\doi\relax


\appendix

\section{Image of the Ellis wormhole}
\label{app:Ellisimage}

\begin{figure}
  \begin{tabular}{cc}
    \begin{minipage}{\hsize}
      \begin{center}
        \includegraphics[width=\linewidth]{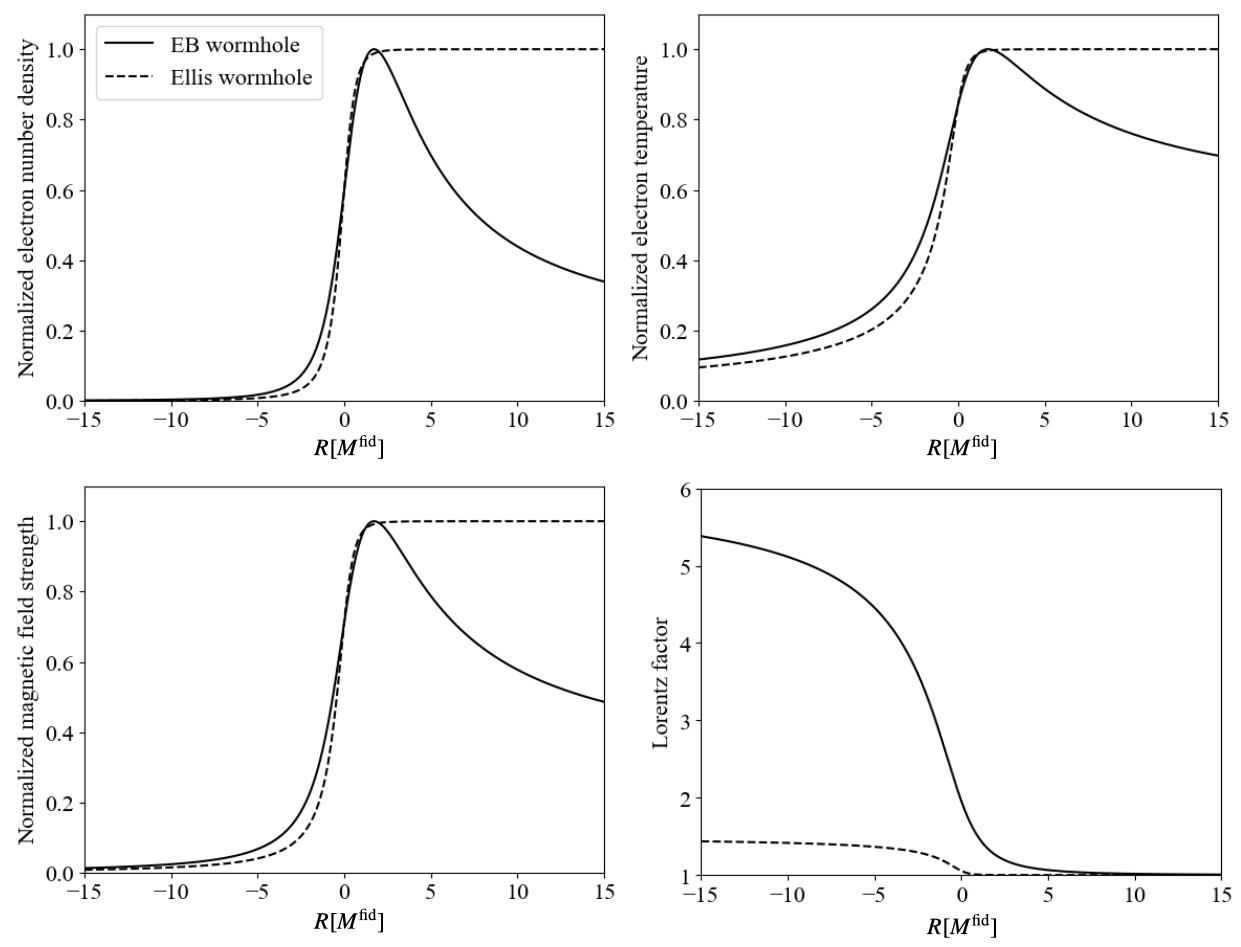}
         \caption{Distributions of the electron number density (top left), the electron temperature (top right), the magnetic field strength (bottom left), and the Lorentz factor (bottom right). 
         The solid and dashed curves correspond to the EB wormhole case and the Ellis wormhole case, respectively. 
         The electron number density, the electron temperature, and the magnetic field strength are normalized by their respective maximum values. 
         The distributions are shown in the range used in the GRRT calculation,
        $-15M^\mathrm{fid} < R < 15M^\mathrm{fid}$, to which the transonic solutions are truncated.}
         \label{fig:gmm_g_mless}
       \end{center}
     \end{minipage}
   \end{tabular}
\end{figure}

To assess the effect of differences in fluid distributions between our study and the previous study~\cite{Ohgami:2015nra}, we calculated the image of the Ellis wormhole. 
Since the Ellis wormhole has $M = 0$, we set $\ell = M^\mathrm{fid}$ and
measure lengths in units of $M^\mathrm{fid}$, as in the fiducial-mass EB
wormhole case. 
In~\cite{Ohgami:2015nra}, the dust fluid in the Ellis wormhole spacetime was considered. 
The steady-state spherical dust solution can be described by a constant velocity solution. 
The authors have denoted the constant velocity as $u_{\infty}$. 
If $u_{\infty}$ vanishes, the number density $n$ becomes an arbitrary function of $R$. 
However, if $u_{\infty} \neq 0$, the number density becomes $n \propto (R^2 + \ell^2)^{-1}$. 
On the other hand, we consider the polytropic fluid with $\Gamma=4/3$, where $\Gamma$ is the adiabatic index. 
As in the EB wormhole case, we obtain the transonic solution numerically in
$-200M^\mathrm{fid} < R < 200M^\mathrm{fid}$ and truncate it to
$-15M^\mathrm{fid} < R < 15M^\mathrm{fid}$ in the GRRT calculation. 
Figure \ref{fig:gmm_g_mless} shows the distributions of the electron number density, the electron temperature, the magnetic field strength, and the Lorentz factor for both the Ellis wormhole and the EB wormhole. 
As seen from the bottom right panel of Fig.~\ref{fig:gmm_g_mless}, in our case, the Lorentz factor of the fluid is not constant for the Ellis wormhole.

\begin{figure}
  \begin{tabular}{cc}
    \begin{minipage}{\hsize}
      \begin{center}
        \includegraphics[width=\linewidth]{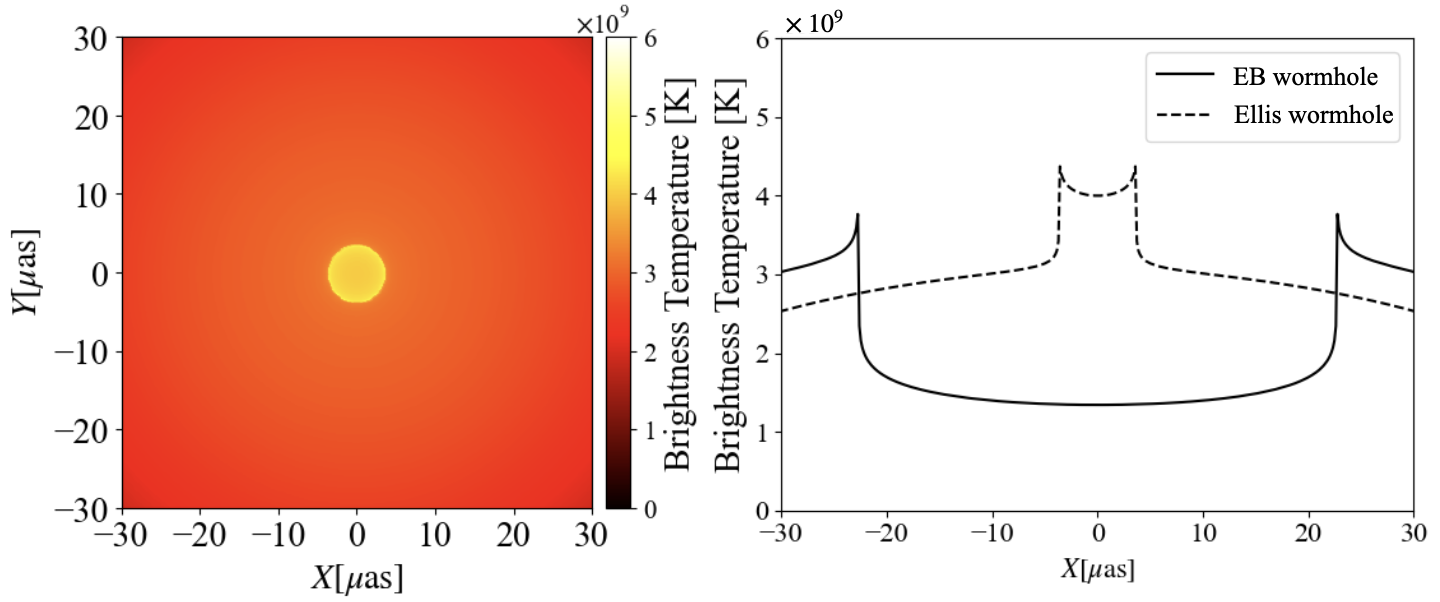}
         \caption{Simulated image of the Ellis wormhole (left panel) and its brightness temperature profile along the $Y=0$ line (right panel). 
         In the right panel, the solid curve is the brightness temperature profile for the EB wormhole, which is the same curve as the solid curve in Fig.~\ref{fig:slice}.}
         \label{fig:image_mless}
       \end{center}
     \end{minipage}
   \end{tabular}
\end{figure}

Figure~\ref{fig:image_mless} shows the simulated image of the Ellis wormhole (left panel) and the brightness temperature profile along the $Y=0$ (right panel). 
For the Ellis wormhole, the region inside the photon ring is brighter than the region outside the photon ring. 
\begin{figure}
  \begin{tabular}{cc}
    \begin{minipage}{\hsize}
      \begin{center}
        \includegraphics[width=\linewidth]{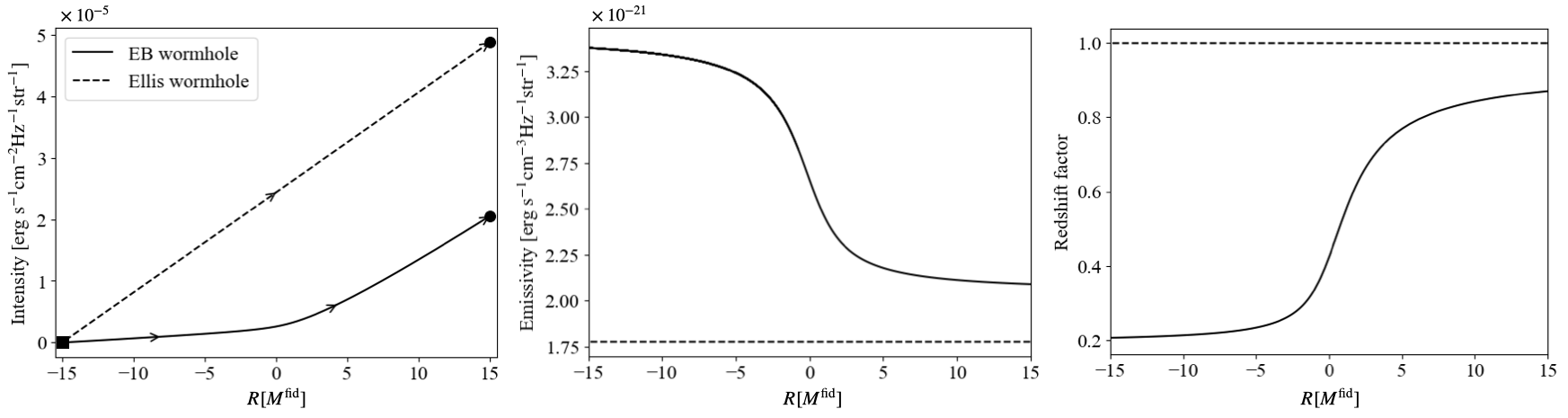}
         \caption{The intensity (left), the emissivity in the fluid rest frame (center), and the redshift factor (right) along the geodesic of a photon that reaches the center of the screen. 
         In the left panel, the black square is the point at which the intensity is set to zero. 
         On the other hand, the black circle is the point at which the geodesic reaches $R = 15M^{\rm fid}$ on its way to the screen; the intensity does not change beyond this point, and hence the value at the black circle corresponds to the intensity at the screen. 
         The arrow indicates the direction of the photon propagation, along which the intensity accumulates.
         }
         \label{fig:center_mless}
       \end{center}
     \end{minipage}
   \end{tabular}
\end{figure}
This is because the redshift factor for the photons reaching the region inside the photon ring is close to unity in the Ellis wormhole spacetime (see the dashed curve of the right panel in Fig.~\ref{fig:center_mless}). 
In the Ellis wormhole, since the lapse function is unity, the redshift factor is primarily determined by the motion of the accreting matter. 
As seen from the bottom right panel of Fig.~\ref{fig:gmm_g_mless}, the Lorentz factor remains small throughout the spacetime in the Ellis wormhole case, and its variation is small, so that the redshift factor is nearly constant. 
Furthermore, the emissivity in the fluid rest frame in the Ellis wormhole case is approximately constant with respect to $R$ (the center panel in Fig.~\ref{fig:center_mless}). 
When both the redshift factor and the emissivity are constant, the intensity is proportional to the path length of the photon. 
Therefore, the intensity for the photon reaching the center of the screen increases monotonically in the Ellis wormhole (the left panel in Fig.~\ref{fig:center_mless}). 
As a result, the region inside the photon ring is brighter than the region outside the photon ring in the Ellis wormhole.

\bibliographystyle{ptephy}
\bibliography{apssamp.bib}

\end{document}